\documentclass[twocolumn,showpacs,preprintnumbers]{revtex4}
\usepackage{amsmath}
\usepackage{amssymb}
\usepackage{graphicx}
\usepackage{dcolumn}
\usepackage{bm}

\begin{document}

\title{Interference of the evolutions of Rb and Na Bose-Einstein condensates}
\author{Ma Luo, Chengguang Bao, Zhibing Li\footnote{%
Corresponding author: stslzb@mail.sysu.edu.cn}}
\affiliation{State Key Laboratory of Optoelectronic Materials and Technologies \\
School of Physics and Engineering \\
Sun Yat-Sen University, Guangzhou, 510275, P.R. China}

\begin{abstract}
Spin evolution of a mixture of two species of spin-1 Bose-Einstein
condensates (Rb and Na) has been investigated beyond the mean field
theory. Both analytical expression and numerical results on the
populations of spin-components are obtained. The interference is
found to depend on the initial states and the inter-species
interaction sensitively. A number of phenomena arising from the
interference have been predicted. In particular, periodic behavior
and the alternate appearances of a zone of oscillation and a long
quiet zone are found.
\end{abstract}

\maketitle

Optical traps of spin-1 to spin-3 bosonic atoms have been
implemented
experimentally, which liberate the freedoms of spin of the atoms \cite%
{ho98,ohmi98,stam98,sten98,goel03,grie05}. At low temperature, the trapped
atoms will form\textbf{\ }a spinor\textbf{\ }Bose-Einstein condensate. The
eigen-states of the system conserve the total spin $S$ and its Z-component $%
M $. If the system is given at an initial state which is not an eigen-state
but a superposition of them, e.g., a Fock-state, the initial state will
evolve afterward. The evolution of spins is an exquisite process because it
depends not only on the interactions but also very sensitively on the
initial states. The study of spin-evolution is not only an attractive topic
but also promising in promoting micro-techniques. This topic has now been
extensively studied experimentally and theoretically. One can measure and
calculate the evolution of the average populations of the three spin
components $\mu =1,0,-1$. This process is called spin mixing or spin
dynamics \cite%
{sore01,duan02,molm94,chang2004,youli2005,law98,pu99,diener2006}. However,
how the spin-evolution of a species would affect the evolution of another
species when they are mixed up in a trap is a topic remains untouched.

In a previous work \cite{Luo2007}, beyond the mean field theory, we have
succeeded in giving a formula governing the spin-evolution of spin-1
Bose-Einstein condensates. On the other hand, the condensation of a mixture
of two species of spin-1 atoms is in principle possible to be realized. A
study of the eigen-states of the mixture has been given in \cite{Luo2006}.
Where the total spin-states are labeled by good quantum numbers $%
(S_{A},S_{B},S,M)$, where $S_{A}$($S_{B}$) is the total spin of a species
A(B), and $S$ and $M$\ are for the whole system. \ As a generalization of
the above two works, in this paper we study the spin-evolution of the
mixture. It is expected that the interference due to the inter-species
interaction will lead to new physics, which might enrich the charming of
this attractive field.

It is assumed that the two species, A and B, contain $N_{A}$ and
$N_{B}$
spin-1 atoms, respectively. The intra-species interaction is $%
V_{ij}^{C}=(c_{0}^{C}+c_{2}^{C}\mathbf{F}_{i}\mathbf{\cdot F}_{j})\delta (%
\mathbf{r}_{i}\mathbf{-r}_{j})$, where $\mathbf{F}_{i}$ is the operator of
the spin of the $i-th$\ atom, $C=A$ or $B$. The inter-species interaction is
$V_{ij}^{AB}=(c_{0}^{AB}+c_{2}^{AB}\mathbf{F}_{i}\mathbf{\cdot F}_{j})\delta
(\mathbf{r}_{i}\mathbf{-r}_{j})$. Under the single mode approximation (SMA)
all the atoms of the same species have the same spatial wave function $\phi
_{A}$ or $\phi _{B}$. We shall focus at the evolution of the spins, while $%
\phi _{A}$ and $\phi _{B}$ are assumed not to be changed with time.
Accordingly, the Hamiltonian responsible for the spin evolution reads
\begin{equation}
H=g_{A}\hat{\mathbf{S}}_{A}^{2}+g_{B}\hat{\mathbf{S}}_{B}^{2}+g_{AB}\hat{%
\mathbf{S}}^{2}  \label{e1}
\end{equation}%
where $\hat{\mathbf{S}}_{A}$, $\hat{\mathbf{S}}_{B}$, and $\hat{\mathbf{S}}$
are the total spin operators of the species A, B, and the whole of them,
respectively.
\begin{eqnarray}
g_{A} &=&\frac{1}{2}(c_{2}^{A}\int d\mathbf{r}|\phi _{A}|^{4}-c_{2}^{AB}\int
d\mathbf{r}|\phi _{A}|^{2}|\phi _{B}|^{2})  \notag \\
g_{B} &=&\frac{1}{2}(c_{2}^{B}\int d\mathbf{r}|\phi _{B}|^{4}-c_{2}^{AB}\int
d\mathbf{r}|\phi _{A}|^{2}|\phi _{B}|^{2})  \label{e2} \\
g_{AB} &=&\frac{1}{2}c_{2}^{AB}\int d\mathbf{r}|\phi _{A}|^{2}|\phi _{B}|^{2}
\notag
\end{eqnarray}%
Accordingly, the eigen-states of the Hamiltonian read
\begin{eqnarray}
|\alpha ,M\rangle &=&|S_{A},S_{B},S,M\rangle  \notag \\
&=&\sum_{M_{A},M_{B}}C_{S_{A}M_{A}S_{B}M_{B}}^{SM}|\vartheta
_{S_{A},M_{A}}^{N_{A}}\rangle |\vartheta _{S_{B},M_{B}}^{N_{B}}\rangle
\label{e3}
\end{eqnarray}%
where $\alpha $ denotes the set $(S_{A},S_{B},S)$, $|\vartheta
_{S_{C},M_{C}}^{N_{C}}\rangle $\ denotes a all-symmetric spin-state of the
C-species with spin $S_{C}$ and its Z-component $M_{C}$. Then $S_{A}$ and $%
S_{B}$ are coupled to the total spin $S$ via the Clebsch-Gordan
coefficients. Evidently, $S_{C}\leq N_{C}$, $|S_{A}-S_{B}|\ \leq S\leq
S_{A}+S_{B}$, and $S\geq M$. Besides, due to the requirement of symmetry, $%
N_{C}-S_{C}$ must be even and $|\vartheta _{S_{C},M_{C}}^{N_{C}}\rangle $ is
unique. \cite{Bao2004b,JK} When $N_{A}$, $N_{B}$, and $M$ are given, the set
of eigen-states with $S_{A}$, $S_{B}$, and $S$ satisfying the above
requirement are complete for the Hamiltonian eq.(\ref{e1}).

Incidentally, we have assumed that the external magnetic field is
exactly zero. In experiments, if a weak residual magnetic field
exists, in order that the effect of the quadratic Zeeman term can be
neglected, from the experience of the single-species evolution, the
field should be weaker than 10$^{-4}Gauss$.

A Fock-state of a species is denoted as $|N_{C1},N_{C0},N_{C,-1}\rangle $,
where $N_{C\mu }$ is the number of C-atoms with the spin-component $\mu $ ($%
0 $ or $\pm 1$), and $N_{C}=\sum_{\mu }N_{C\mu }$. \ Initially, it is
assumed that each species is in a Fock-state. Thus the initial state $%
|O\rangle =|N_{A1}^{I},N_{A0}^{I},N_{A,-1}^{I}\rangle
|N_{B1}^{I},N_{B0}^{I},N_{B,-1}^{I}\rangle ,$ where $N_{C\mu }^{I}$ are the
initial values. \ The associated formal time-dependent solution is just $%
|\Psi (t)\rangle =e^{-iHt/\hbar }|O\rangle $. When $|O\rangle $ is expanded
by the eigen-states of $H$\ (eq.(\ref{e3})), the formal solution can be
written in a practicable form as\
\begin{equation}
|\Psi (t)\rangle =\sum_{\alpha }e^{-iE_{\alpha }t/\hbar }|\alpha ,M\rangle
\langle \alpha ,M|O\rangle  \label{ee3}
\end{equation}%
where $M=N_{A1}^{I}-N_{A,-1}^{I}+N_{B1}^{I}-N_{B,-1}^{I}$, $E_{\alpha
}=g_{A}S_{A}(S_{A}+1)+g_{B}S_{B}(S_{B}+1)+g_{AB}S(S+1)$. Since the set $%
|\alpha \rangle $ is complete, the above expression is an exact solution of
the Hamiltonian if the coefficients $\langle \alpha ,M|O\rangle $ can be
exactly calculated. \ This is given in the appendix. Since $M$ is conserved
during the evolution, $|\alpha ,M\rangle $ is written simply as $|\alpha
\rangle $ in the follows.

Our aim is to calculate the time-dependence of the populations
\begin{equation}
\langle \Psi (t)|\hat{N}_{\mu }^{C}|\Psi (t)\rangle /N_{C}\equiv P_{\mu
}^{C}(t)  \label{e4}
\end{equation}%
where $\hat{N}_{\mu }^{C}$ is the operator of number of C-atoms in $\mu $. \
Obviously, $\sum_{\mu }P_{\mu }^{C}(t)=1$ holds at any time. Inserting (\ref%
{ee3}) into (\ref{e4}), we have
\begin{equation}
P_{\mu }^{C}(t)=B_{\mu }^{C}+O_{\mu }^{C}(t)  \label{e5}
\end{equation}%
with
\begin{equation}
B_{\mu }^{C}=\frac{1}{N_{C}}\sum_{\alpha }\langle \alpha |\hat{N}_{\mu
}^{C}|\alpha \rangle \langle O|\alpha \rangle ^{2}  \label{e6}
\end{equation}%
\begin{equation}
O_{\mu }^{C}(t)=\frac{2}{N_{C}}\underset{\alpha <\alpha ^{\prime }}{\sum }%
\cos [(E_{\alpha ^{\prime }}-E_{\alpha })\frac{t}{\hbar }]\langle \alpha
^{\prime }|\hat{N}_{\mu }^{C}|\alpha \rangle \langle O|\alpha \rangle
\langle O|\alpha ^{\prime }\rangle  \label{e7}
\end{equation}%
The expression of $\langle \alpha ^{\prime }|\hat{N}_{\mu }^{C}|\alpha
\rangle $ is also given in the appendix. The above formulae are a direct
generalization of those given in eq.(3) to (7) of \cite{Luo2007} for a
single species, and they reveal an oscillation described by $O_{\mu }^{C}(t)$
around a background $B_{\mu }^{C}$. \textit{It is noticeable that }$B_{\mu
}^{C}$\textit{\ depends not at all on the interactions, but completely
determined by the inherent symmetry and the initial state. In fact, the
dynamics affects only the factor }$E_{\alpha ^{\prime }}-E_{\alpha }$\textit{%
\ in }$O_{\mu }^{C}(t)$ (\textit{this would cause a re-scale of time as we
shall see}).\textit{\ All other factors in (7) and (8) are independent of
dynamics.} Furthermore, from eq.(\ref{e7}) and the appendix, $\langle \alpha
^{\prime }|\hat{N}_{\mu }^{A}|\alpha \rangle $ is nonzero only if $%
S_{B}^{\prime }=S_{B}$, it implies that the associated $E_{\alpha ^{\prime
}}-E_{\alpha }$\ does not depend on $g_{B}$, and therefore $P_{\mu }^{A}$\
does not depend on $g_{B}$. Similarly, $P_{\mu }^{B}$\ does not depend on $%
g_{A.}$.

We choose $^{87}$Rb as A and $^{23}$Na as B. They are confined by a
parabolic potential with frequency $\omega =300/\sec $ (unless specified).
The parameters for the inter-species interaction are from Table III of ref.
\cite{Pashov} . The set of coupled Gross-Pitaevskii equations for the
mixture derived in our previous paper \cite{Luo2006} has been used to obtain
the spatial wave functions. Numerical results on $P_{\mu }^{C}(t)$ for
selected cases are given in the follows.

\bigskip

(I) Interference of two initially fully polarized systems

If both species are polarized and they have the same direction of
polarization, no spin-flips would occur due to the conservation of the total
magnetization. Therefore only the cases with opposite directions are
considered. Let an initial state be hereafter labeled as $%
(N_{A1}^{I},N_{A0}^{I},N_{A,-1}^{I})_{A}(N_{B1}^{I},N_{B0}^{I},N_{B,-1}^{I})_{B}
$. For the case $(N_{A},0,0)_{A}(0,0,N_{B})_{B}$, namely, all
A-atoms are up and all B-atoms are down initially,\ the factors in
eq.(\ref{e7}) have simpler forms. They are
\begin{equation}
\langle O|\alpha \rangle =\delta _{S_{A},N_{A}}\delta
_{S_{B},N_{B}}\delta
_{M,N_{A}-N_{B}}C_{N_{A},N_{A},N_{B},-N_{B}}^{S,N_{A}-N_{B}}
\label{e7a}
\end{equation}%
and
\begin{eqnarray}
& &\langle \alpha ^{\prime }|\hat{N}_{\mu }^{A}|\alpha \rangle
\nonumber \\&=&\sum_{K}C_{N_{A},K,N_{B},N_{A}-N_{B}-K}^{S^{\prime
},N_{A}-N_{B}}C_{N_{A},K,N_{B},N_{A}-N_{B}-K}^{S,N_{A}-N_{B}}
\nonumber
\\& &\cdot [A(N_{A},N_{A},K,\mu )^{2}+B(N_{A},N_{A},K,\mu )^{2}]
\label{e7b}
\end{eqnarray}
where the quantities inside the brackets are the fractional
parentage coefficients given in the appendix \cite{Bao2004a}. Since
both the Clebsch-Gordan and fractional parentage coefficients arise
purely from symmetry, it is clear that symmetry is decisive to the
evolution.

Due to eq.(\ref{e7a}) the factor $\langle O|\alpha \rangle \langle
O|\alpha ^{\prime }\rangle $ in eq.(\ref{e7}) assures
$S_{A}=S_{A}^{\prime }=N_{A}$, and $S_{B}^{\prime }=S_{B}=N_{B}$ . \
Accordingly, $E_{\alpha ^{\prime }}-E_{\alpha }=g_{AB}[S^{\prime
}(S^{\prime }+1)-S(S+1)]$. It implies three points: (i) The
evolutions of $P_{\mu }^{A}(t)$ and $P_{\mu }^{B}(t)$ do not depend
on $g_{A}$ and $g_{B}$, but on $g_{AB}\propto c_{2}^{AB}$. (ii) The
effect of interactions is embodied uniquely via the factor
$g_{AB}t$. Therefore the effect is just a re-scale of time, namely,
to accelerate or slow down the evolution. The general features of
evolutions are determined by inherent symmetry and the initial
conditions. (iii) Since $S^{\prime }(S^{\prime }+1)-S(S+1)$ must be
an even integer $I_{e}$, the time-dependent factor in eq.(\ref{e7})
can be rewritten as $\cos (I_{e}g_{AB}t/\hbar )$.\ Thus $P_{\mu
}^{A}(t)$ and $P_{\mu }^{B}(t)$ are strictly periodic with the same
period $t_{p}\equiv \pi \hbar /|g_{AB}|$, and they are symmetric
with respect to $t_{p}/2$. So, if the period can be determined, it
provides an opportunity to measure the A-B interaction. \ In what
follows, $t_{p}$\ is used as the unit of time.

\begin{figure}[tbp]
\scalebox{1}{\includegraphics{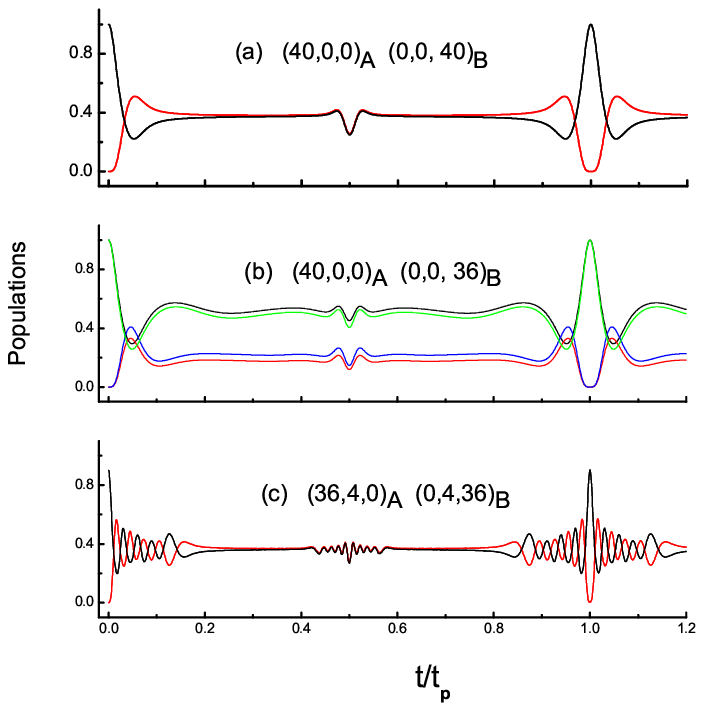}}
\caption{Interference of the evolutions of two initially fully polarized
condensates A(Rb) and B(Na). $\mathbf{P}_{1}^{A}(t)$ (black), $\mathbf{P}%
_{-1}^{A}(t)$ (red), $\mathbf{P}_{1}^{B}(t)$ (blue), and $\mathbf{P}%
_{-1}^{B}(t)$ (green) are plotted against $\protect\tau =t/t_{p}$. The
initial particle numbers of the spin-components $%
(N_{C1},N_{C0},N_{C,-1})_{C} $ ($C=A$ or $B$) are marked in each panel. $%
\protect\omega =300/\sec $ is assumed. The implication of the colors and the
magnitude of $\protect\omega $ are the same in the follows. In (a) and (c)
the black curve overlaps the green, the red overlaps the blue. }
\label{LF1}
\end{figure}

When the initial state is $(40,0,0)_{A}(0,0,40)_{B}$, $P_{\pm
1}^{C}(t)$ are plotted in Fig.\ref{LF1}a against $\tau \equiv
t/t_{p}$ from 0 to 1.2. Where the strict periodicity together with
the symmetry with respect to $\tau =1/2$ are clearly shown. If the
inter-species interaction is removed, there would be no spin-flips
due to the conservation of the magnetization of each species. Due to
the interaction, in the early stage, there is a very strong process
of spin-flips initiated by the collisions between the up A-atoms and
down B-atoms. Thereby $P_{1}^{A}$ drops rapidly from 1 to a minimum
$0.222$ at $\tau =\tau _{\min }=0.052$, while $P_{-1}^{A}$ and
$P_{0}^{A}$ increase from zero rapidly. When $\tau \approx 0.1$, the
system arrives at a steady stage with $P_{1}^{A}=P_{-1}^{A}\approx
3/8$ and $P_{0}^{A}\approx 1/4.$ The equality
$P_{1}^{A}=P_{-1}^{A}$\ implies that the system has given up all its
previous polarization. Afterward, the system keeps quiet in a long
duration until $\tau \approx 0.9$. Then, in the last stage of the
period, the previous magnetization is completely recovered. Thus, a
\textit{zone of oscillation (ZoS)} followed by a \textit{quiet zone,
}and again a\textit{\ ZoS} appear successively and repeatedly.
However, there is a small turbulence occurring at the right middle
of the quiet zone.

During all the time the behavior of the B-atoms match exactly with the
A-atoms, namely, $P_{\mu }^{A}(\tau )=P_{-\mu }^{B}(\tau )$ due to symmetry
as expected. It is emphasized that the appearance of the quiet zone is a
noticeable and quite popular feature of spin-evolution. This feature is in
nature a quantum phenomenon of interference as shown by eqs. (\ref{e7a}), (%
\ref{e7b}), and (\ref{e7}). From these formulae it is clear that the effect
of interactions is simply to adjust the length of the zone (longer or
shorter).

It is noted that $g_{AB}$ $\propto \int d\mathbf{r}|\phi _{A}|^{2}|\phi
_{B}|^{2}$ , which is roughly proportional to $\omega ^{6/5}N^{-3/5}$. \cite%
{NW} Thus it is clear that increasing the confinement or reducing the
particle number will accelerate the evolution and shorten the period $t_{p}$
(e.g., Numerically, when $N_{A}=N_{B}=40$ and $\omega =300$, we have $%
t_{p}=558.6\sec $. If $\omega $\ increases by ten times, $t_{p}$=35.2$\sec $%
). In eq.(\ref{e7}) all factors, except the time-dependent factor, are
irrelevant to $\omega $ and insensitive to $N$ (if the ratios of the $%
N_{C\mu }^{I}$\ of the initial state remain the same). Therefore the
change of $\omega $\ is simply equivalent to a re-scale of time, and
the variation of $N$ (in the above sense) would cause only a slight
change quantitatively but not qualitatively.

In order to see the effect of the asymmetry of particle numbers, the initial
case $(40,0,0)_{A}(0,0,36)_{B}$ is considered as shown in Fig.\ref{LF1}b. A
remarkable feature is the great separation of $P_{1}^{C}$ and $P_{-1}^{C}$
curves in the quiet zone. It implies that the initial polarizations of both
species are partially preserved.

Experimentally, the polarization might not be perfect. An example is given
in Fig.\ref{LF1}c, where each species has four atoms in $\mu =0$ initially.
Comparing \ref{LF1}c with \ref{LF1}a, it is clear that the four $\mu =0$
atoms of each species cause additionally four rounds of oscillations in each
ZoS.\ Thus the ZoS becomes longer and the quiet zone becomes shorter
accordingly. In each round of oscillation $P_{1}^{C}$ and $P_{-1}^{C}$ keep
to have reverse phases. In general, when the initial state is $%
(N_{A}-K,K,0)_{A}(0,K,N_{A}-K)_{B}$, it was found that $K$\ rounds of
oscillation will emerge in each ZoS additionally.

\bigskip

(II) The interference of a fully polarized system with a zero-polarized
system

\begin{figure}[tbp]
\scalebox{1}{\includegraphics{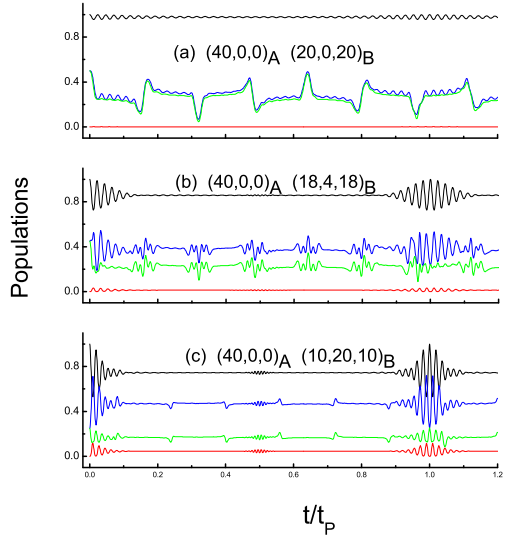}}
\caption{Interference of an initially fully polarized condensate with an
initially non-polarized condensate.}
\label{LF2}
\end{figure}

When the initial state is (40,0,0)$_{A}$(20,0,20)$_{B}$, the evolution is
shown in Fig.\ref{LF2}a which is greatly different from Fig.\ref{LF1}a.
Meanwhile a half of B-atoms are up while the other half are down initially.
Thus the total spin of the group of up-B-atoms and that of the down-B-atoms
have the same magnitude but opposite directions. Therefore the vector sum of
them must be small. Hence, those $|\alpha \rangle $\ with a large $S_{B}$
will lead to a very small $\langle \alpha |O\rangle $, and therefore can be
omitted from the expansion of $\Psi (t)$ (refer to eq.(\ref{ee3})). In other
words, during the evolution, $S_{B}$ remains to be small. Consequently, the
magnetization of the B-atoms, $M_{B}$, remains to be small. Thus, in order
to keep $M=M_{A}+M_{B}$\ to be conserved, $P_{1}^{A}(\tau )$ must be always
close to 1. This deduction is confirmed by Fig.\ref{LF2}a.

Since the A-atoms are always close to be fully-polarized, they are inert.
One might therefore guess that the B-atoms would behave just as if the
A-atoms do not exist. This suggestion is partially true. Comparing the
curves of $P_{\mu }^{B}(\tau )$ shown in Fig.\ref{LF2}a\ with those with the
A-atoms removed (not plotted), they are one-to-one similar in pattern, but
different in the scale of time. In fact, the former is a compressed version
of the latter, namely, the evolution of the former is more rapid. For an
example, the former is roughly periodic with a period $\approx $0.64$t_{p}$,
while the latter is exactly periodic with a period 1.65$t_{p}$. It is
mentioned that the Hamiltonian of the latter is $g_{B}^{free}\hat{\mathbf{S}}%
_{B}^{2}$, where $g_{B}^{free}=\pi \hbar /(\frac{1}{2}c_{2}^{B}\int dr|\phi
_{B}|^{4})$. On the other hand, both $c_{2}^{B}$ and $c_{2}^{AB}$ are
contributed to $g_{B}$. It turns out that these two strengths have opposite
signs. This leads to $g_{B}\gg g_{B}^{free}$ (refer to eq.(\ref{e2})), and
therefore a faster evolution. Incidentally, $P_{1}^{B}(\tau )$ plotted in
Fig.\ref{LF2}a overlaps nearly with $P_{-1}^{B}(\tau )$. This confirms the
previous deduction that $M_{B}$ remains small.

On the other hand, although $P_{\mu }^{A}(\tau )$\ varies very slightly with
time, it is strictly periodic. In general, when one of the species (say, A)
is fully-polarized initially, its evolution is strictly periodic\
disregarding how B\ is initially. This arises because meanwhile the factor $%
\langle O|\alpha \rangle \langle O|\alpha ^{\prime }\rangle $ in
eq.(\ref{e7} ) assures $S_{A}=S_{A}^{\prime }=N_{A}$, while the
factor $\langle \alpha ^{\prime }|\hat{N}_{\mu }^{A}|\alpha \rangle
$ assures $S_{B}^{\prime }=S_{B} $ (refer to the appendix). \
Accordingly, the time-dependent factor in (\ref{e7}) becomes $\cos
([S^{\prime }(S^{\prime }+1)-S(S+1)]g_{AB}t/\hbar )=\cos
(I_{e}g_{AB}t/\hbar )$, thus the strict periodicity with the period
$t_{p}$ is clear.

Incidentally, if A and B\ interchange their initial status, namely, the
initial state is (20,0,20)$_{A}$(40,0,0)$_{B}$, then the B-atoms would be
inert, while the behavior of $P_{\mu }^{A}$\ would be similar to the $P_{\mu
}^{B}$ of Fig.\ref{LF2}a in pattern but with a longer period. The reason is
that both $c_{2}^{A}$ and $c_{2}^{AB}$ are negative resulting in a weaker $%
g_{A}$ and therefore a slower evolution.

When the initial state of Fig.\ref{LF2}a is slightly changed to (40,0,0)$%
_{A} $(18,4,18)$_{B}$, namely, a few B-atoms are initially in $\mu =0$, the
evolution is thereby changed dramatically as shown in \ref{LF2}b. Meanwhile,
the constraint caused by $\langle \alpha |O\rangle $\ is not so strict and
those $|\alpha \rangle $ with a larger $S_{B}$\ can be contained in $\Psi
(t) $,\ so that a greater part of the total magnetization can be transferred
to the B-atoms. Accordingly, $P_{1}^{A}(\tau )$ may differ from 1
explicitly, and the interference of the two species becomes stronger. In \ref%
{LF2}b, the periodicity of the A-species with the period $t_{p}$ is clearly
shown, and the two ZoS together with the quiet zone are also found. In the
quiet zone of $P_{\mu }^{A}$, we have $P_{1}^{A}-P_{-1}^{A}\approx 0.85$. It
implies that 15\% of magnetization has been transferred to B.\ However, $%
P_{-1}^{A}$ remains to be very small. It is noticeable that $P_{\mu }^{B}$
is strongly affected by $P_{1}^{A}$. In particular, when the A-atoms
oscillate strongly, the B-atoms also. Thus, $P_{\mu }^{B}$ seems to be a
mixture of two modes, one matches the evolution of A with the two ZoS, the
other one is just the residual of its original mode.

When half of the B-atoms have $\mu =0$\ initially, the evolution is plotted %
\ref{LF2}c which is similar to \ref{LF2}b. However, $P_{\mu }^{B}$ and $%
P_{\mu }^{A}$ match more strongly with each other in reverse phase (a peak
of $P_{\mu }^{B}$\ matches a dip of $P_{\mu }^{A}$), and the residual mode
in $P_{\mu }^{B}$ becomes very weak. Thus \ref{LF2}a and \ref{LF2}c are two
typical examples to show a very weak and a very strong interferences,
respectively.

\bigskip

(III) Effect of the inter-species interaction

\begin{figure}[tbp]
\scalebox{1}{\includegraphics{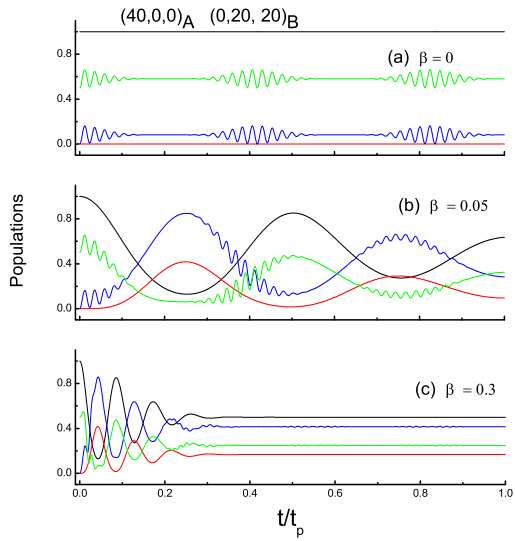}}
\caption{ Interference of an initially fully polarized condensate with a
half polarized condensate. The actual inter-species interaction has been
reduced by a factor $\protect\beta $ which is marked in each panel. }
\label{LF5}
\end{figure}

Let the realistic $V_{ij}^{AB}$ be changed to $\beta V_{ij}^{AB}$, then we
study the effect of the adjustable strength $\beta $. When the initial
state, for an example, is $(40,0,0)_{A}(0,20,20)_{B}$, three versions with $%
\beta =0,\ 0.05$ \ and 0.3, respectively, are plotted in Fig.\ref{LF5}a to
c. It is reminded that, once the inter-species interaction is removed, the
evolution of each species is strictly periodic with the period $%
t_{p}^{C}=\pi \hbar /(\frac{1}{2}c_{2}^{C}\int d\mathbf{r}|\phi _{C}|^{4})$.
\cite{Luo2007} In particular, $t_{p}^{A}\ =3.21t_{p}$ and $t_{p}^{B}\
=1.65t_{p}$ (with our parameters). In Fig.\ref{LF5}a with $\beta =0$, the
A-atoms do not evolve, while the B-atoms evolve according to the law given
in ref.\cite{Luo2007}. In particular, each time when $t$\ is close to $%
Kt_{p}^{B}/4$, where $K=0,1,2,\cdot \cdot \cdot $ $,$ the ZoS appears
wherein $P_{\mu }^{B}(\tau )$\ would contain a few rounds of oscillation.
Between two adjacent ZoS, it is the quiet zone.

The case with $\beta =0.05$ is shown in \ref{LF5}b. Two points are
noticeable. (i) Since the A-atoms have been fully-polarized initially, if
they can evolve via the help of the B-atoms, they should behave exactly
periodically as mentioned previously. The period should be $t_{p}^{(\beta
)}\equiv \pi \hbar /|g_{AB}^{\beta }|$, where $g_{AB}^{\beta }$ is similarly
defined as $g_{AB}$ but with $c_{2}^{AB}$\ replaced by $\beta c_{2}^{AB}$.
Since $|g_{AB}^{\beta }|\ll |g_{AB}|$ , $t_{p}^{(\beta )}$ is very large. In
fact, the range of $\tau $\ in \ref{LF5}b is only a small portion of $%
t_{p}^{(\beta )}$. (ii) Although $\beta V_{ij}^{AB}$ is so weak, $P_{\mu
}^{B}$ is seriously affected by the A-atoms. Consequently, $P_{\mu }^{B}$ is
a mixture of two modes, just as the previous case in Fig.\ref{LF2},\ one
matches $P_{\mu }^{A}$\ in reverse phase, while the other is the residual of
the original mode. Thus, both systems oscillate with a large amplitude and a
very low frequency, while additional small oscillation takes place in $%
P_{\mu }^{B}$ occasionally due to the second mode.

When $\beta $\ increases further, $t_{p}^{(\beta )}$ becomes shorter, and
the corresponding $P_{\mu }^{C}$ is plotted in Fig.\ref{LF5}c which is a
compression version of \ref{LF5}b towards left. In \ref{LF5}c the range of $%
t $ is close to $t_{p}^{(\beta )}/3$, and we can see the broad quiet zones.
When $\beta =1$, the patterns in \ref{LF5}c are further compressed.

\begin{figure}[tbp]
\scalebox{1}{\includegraphics{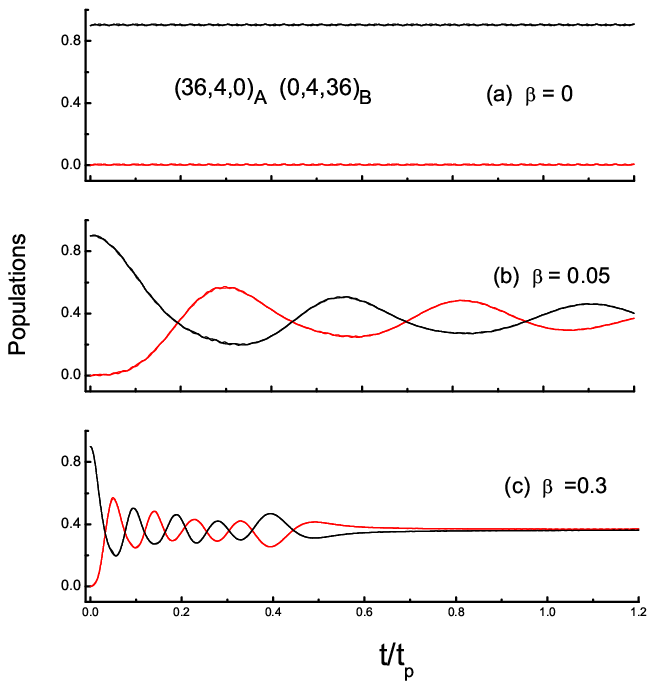}}
\caption{ Interference of two initially polarized condensates. The
polarizations of both systems are not perfect, each contains four $\protect%
\mu =0$ atoms. The actual inter-species interaction has been reduced by a
factor $\protect\beta $ which is marked in each panel.}
\label{LF6}
\end{figure}

One more example is shown in Fig.\ref{LF6} with the initial state $%
(36,4,0)_{A}(0,4,36)_{B}$ corresponding to Fig.\ref{LF1}c. If the
inter-species interaction is removed, both species are inert as shown in \ref%
{LF6}a because they are nearly fully-polarized initially. When $\beta =0.05$
as shown in \ref{LF6}b, the evolutions of both systems become nearly
periodic with the same long period close to $t_{p}^{(\beta )}=\pi \hbar
/|g_{AB}^{\beta }|$ as mentioned before. When $\beta $ is larger, the
pattern is compressed towards left as shown in \ref{LF6}c. A further
compression of \ref{LF6}c leads to Fig.\ref{LF1}c.

\bigskip

In summary, a theoretical tool (the fractional parentage coefficients) has
been introduced to study the spin-evolution of a mixture of the condensates
of $^{87}$Rb (A) and $^{23}$Na (B) atoms. Making use of the
single-mode-approximation, a formula has been derived to describe the
evolution beyond the mean field theory. The evolution of each spin component
$P_{\mu }^{C}(\tau )$ appears as an oscillation around a background. The
background is determined by the inherent symmetry and the initial state, and
is not at all affected by the interactions.  Based on the formula, selected
cases have been studied numerically. A number of predictions on the
character of evolution have been made.

When both species are (nearly) fully polarized in reverse direction
initially (Fig.\ref{LF1}),  this is a case of strong interference.
The evolution of two species are closely match with each other
($P_{\mu }^{A}(\tau )=P_{-\mu }^{B}(\tau )$) and evolve with the
same mode (in reverse phase) and the same period $t_{p}$\ determined
by the inter-species interaction. ZoS and quiet zone appear
alternately. In the former, several rounds of oscillation will
emerge in $P_{\mu }^{C}(\tau )$, and the total magnetization of each
species will undergo a great change (from being fully polarized to
zero polarized, or vice versa). The appearance of quiet zones is
quite popular in spin-evolution and is a quantum phenomenon of
interference.

When A is fully polarized while B is zero-polarized, and if B does not have $%
\mu =0$ atoms initially (Fig..\ref{LF2}a), this is a case of weak
interference. However, if B does have a few $\mu =0$ atoms initially, the
few atoms can work as a catalyst and will cause a strong interference
(comparing \ref{LF2}a and \ref{LF2}b). The sensitivity to the initial $\mu =0
$ atoms is a noticeable point.

When A is fully polarized, disregarding how B is initially, if we introduce $%
\beta $\ to adjust the strength of the inter-species interaction, we
found that the interference can be initiated by a very small $\beta
$ and a new mode of evolution is thereby caused (Fig.\ref{LF5}b).
Accordingly, A evolves with the new mode and B evolves with a
mixture of the new mode and a residual. However, if $\beta $\ is
very small, the new mode will have a very long period, therefore the
interference is difficult to be observed in the early stage.
Nonetheless, when $\beta $\ increases, the period of the new mode
becomes shorter and causes a compression of $P_{\mu }^{C}(\tau )$
towards left (\ref{LF5}c). Finally, both species are dominated by
the new mode.

The above predictions remain to be confirmed experimentally.
Furthermore, the effectiveness of the single-mode approximation and
the particle numbers (if they are very large) deserved to be further
studied.

\bigskip

\begin{acknowledgments}
We appreciate the support from the NSFC under the grants 10574163 and
10674182.
\end{acknowledgments}

\section*{Appendix}

1, The calculation of $\langle O|\alpha \rangle $

The initial state
\begin{equation}
|O\rangle =|N_{A1}^{I},N_{A0}^{I},N_{A,-1}^{I}\rangle
|N_{B1}^{I},N_{B0}^{I},N_{B,-1}^{I}\rangle
\end{equation}
Let
\begin{equation}
M_{C}^{I}=N_{C1}^{I}-N_{C,-1}^{I}
\end{equation}
where $C=A$ or $B$, and $M=M_{A}^{I}+M_{B}^{I}$ . Thus
\begin{equation}
\langle O|\alpha \rangle =C_{S_{A}M_{A}^{I}\
S_{B}M_{B}^{I}}^{SM}D_{M_{A}^{I},N_{A0}^{I}}^{N_{A}S_{A}}D_{M_{B}^{I},N_{B0}^{I}}^{N_{B}S_{B}}
\end{equation}
where $D_{M_{C}^{I},N_{C,0}^{I}}^{N_{C},S_{C}}\equiv \langle
N_{C1}^{I},N_{C0}^{I},N_{C,-1}^{I}|\vartheta _{S_{C},M_{C\
0}}^{N_{C}}\rangle $ is for a single species and is very useful. These
coefficients can be calculated via the following recursion formulae
\begin{eqnarray}
&&\sqrt{(N-N_{0}+M)/(2N)}D_{M,N_{0}}^{N,S} \\
&=&A(N,S,M,1)D_{M-1,N_{0}}^{N-1,S+1}+B(N,S,M,1)D_{M-1,N_{0}}^{N-1,S-1}
\notag  \label{a1}
\end{eqnarray}%
\begin{eqnarray}
&&\sqrt{N_{0}/N}D_{M,N_{0}}^{N,S} \\
&=&A(N,S,M,0)D_{M,N_{0}-1}^{N-1,S+1}+B(N,S,M,0)D_{M,N_{0}-1}^{N-1,S-1}
\notag  \label{a2}
\end{eqnarray}%
\begin{eqnarray}
&&\sqrt{(N-N_{0}-M)/(2N)}D_{M,N_{0}}^{N,S} \\
&=&A(N,S,M,-1)D_{M+1,N_{0}}^{N-1,S+1}+B(N,S,M,-1)D_{M+1,N_{0}}^{N-1,S-1}
\notag  \label{a3}
\end{eqnarray}%
where
\begin{eqnarray}
&&A(N,S,M,\mu ) \\
&=&\left[ \frac{(1+(-1)^{N-S})(N-S)(S+1)}{(2N(2S+1))}\right]
^{1/2}C_{S+1,M-\mu ,\ 1,\mu }^{S,M}  \notag  \label{a4}
\end{eqnarray}%
\begin{eqnarray}
&&B(N,S,M,\mu ) \\
&=&\left[ \frac{(1+(-1)^{N-S})S(N+S+1)}{(2N(2S+1))}\right]
^{1/2}C_{S-1,M-\mu ,\ 1,\mu }^{S,M}  \notag  \label{a5}
\end{eqnarray}%
The above recursion formulae are derived based on the analytical expressions
of the fractional parentage coefficients given in \cite{Bao2004a} and \cite%
{Bao2004b}.

2, The calculation of $\langle \alpha ^{\prime }|\hat{N}_{\mu }^{C}|\alpha
\rangle $

Let C=A, then
\begin{eqnarray}
& &\langle \alpha ^{\prime }|\hat{N}_{\mu }^{A}|\alpha \rangle  \notag \\
&=&\langle S_{A}^{\prime },S_{B}^{\prime },S^{\prime },M|\hat{N}_{\mu
}^{A}|S_{A},S_{B},S,M\rangle \\
&=&\delta _{S_{B}^{\prime }S_{B}}\sum_{M_{A},M_{B}}C_{S_{A}^{\prime
}M_{A},S_{B}M_{B}}^{S^{\prime },M}C_{S_{A}M_{A},S_{B}M_{B}}^{S,M}  \notag \\
&~& ~~~~ \cdot \langle \vartheta _{S_{A}^{\prime },M_{A}}^{N_{A}}|\hat{%
\mathbf{N}}_{\mu }^{A}|\vartheta _{S_{A},M_{A}}^{N_{A}}\rangle  \notag
\end{eqnarray}%
Where the last factor concerns only the A-species, we have
\begin{eqnarray}
&&\langle \vartheta _{S_{A}^{\prime },M_{A}^{\prime }}^{N_{A}}|\hat{N}_{\mu
}^{A}|\vartheta _{S_{A},M_{A}}^{N_{A}}\rangle  \notag \\
&=&\delta _{M_{A}^{\prime },M_{A}}N_{A} \\
&&\{\delta _{S_{A}^{\prime }S_{A}}[(A(N_{A},S_{A},M_{A},\mu
))^{2}+(B(N_{A},S_{A},M_{A},\mu ))^{2}]  \notag \\
&&+\delta _{S_{A}^{\prime },S_{A}-2}A(N_{A},S_{A}-2,M_{A},\mu
)B(N_{A},S_{A},M_{A},\mu )  \notag  \label{a7} \\
&&+\delta _{S_{A}^{\prime },S_{A}+2}A(N_{A},S_{A},M_{A},\mu
)B(N_{A},S_{A}+2,M_{A},\mu )\ \}  \notag
\end{eqnarray}

The derivation of (\ref{a7}) is also based on the fractional parentage
coefficients. For the case C=B, the calculation is similar.

In particular, when $\alpha ^{\prime }=\alpha $, $\mu =0$, and $N_{A}$ is
large, we have
\begin{eqnarray}
\langle \alpha |\hat{N}_{0}^{A}|\alpha \rangle &\approx &\frac{1}{2}%
N_{A}[1-\sum_{M_{A},M_{B}}(C_{S_{A}M_{A},S_{B}M_{B}}^{S,M}M_{A}/S_{A})^{2}]
\notag \\
&\leq &N_{A}/2  \label{a8}
\end{eqnarray}

\clearpage

\begin{thebibliography}{99}
\bibitem{ho98} Tin-Lun Ho, Phys. Rev. Lett., 81, 742 (1998)

\bibitem{ohmi98} T. Ohmi and K. Machida, J. Phys. Soc. Jpn. 67, 1822 (1998)

\bibitem{stam98} D.M. Stamper-Kurn, M.R. Andrews, A.P. Chikkatur, S. Inouye,
H.-J. Miesner, J. Stenger, and W. Ketterle, Phys. Rev. Lett., 80, 2027(1998)

\bibitem{sten98} J. Stenger, S. Inouye, D. M. Stamper-Kurn, H. -J. Miesner,
A. P. Chikkatur, and W. Ketterle, Nature (London) 396, 345 (1998).

\bibitem{goel03} A. Goelitz, T. L. Gustavson, A. E. Leanhardt, R. Low, A. P.
Chikkatur, S. Gupta, S. Inouye, D. E. Pritchard, and W. Ketterle, Phys. Rev.
Lett., 90, 090401 (2003)

\bibitem{grie05} A. Griesmaier, J. Werner, S. Hensler, J. Stuhler, and T.
Pfau, Phys. Rev. Lett., 94, 160401(2005)

\bibitem{sore01} A. Sorensen, L.-M. Duan, J.I. Cirac, and P. Zoller, Nature
409, 63 (2001)

\bibitem{duan02} L.-M. Duan, J.I. Cirac, and P. Zoller, Phys. Rev. A 65,
033619(2002)

\bibitem{molm94} K. Molmer and P. Zoller, Phys. Rev. A50, 67(1994)

\bibitem{chang2004} M.-S. Chang, C.D. Hamley, M.D. Barrett, J.A. Sauer, K.M.
Fortier, W.Zhang, L. You, and M.S. Chapman, Phys. Rev. Lett., 92,
140403(2004)

\bibitem{youli2005} M.S. Chang, Q. Qin, W.X. Zhang, L. You, and M.S.
Chapman, Nature Physics (London) 1, 111(2005)

\bibitem{law98} C.K. Law, H. Pu, and N.P. Bigelow, Phys. Rev. Lett., 81,
5257(1998)

\bibitem{pu99} H. Pu, C.K. Law, S. Raghavan, J.H. Eberly, and N.P. Bigelow,
Rhys. Rev. A., 60, 1463(1999)

\bibitem{diener2006} R.B. Diener and T.L. Ho, arxiv:cond-mat/0608732 (2006)

\bibitem{Luo2007} M. Luo, C.G. Bao, Z.B. Li, Phys. Rev. A77, 043625 (2008)

\bibitem{Bao2004a} C.G. Bao, Acta Scientiarum Naturalium Universitatis
Sunyatseni, \ 43, \ 70 (2004)

\bibitem{Bao2004b} C.G. Bao, Z.B. Li, Phys. Rev. A70, 043620(2004)

\bibitem{JK} J. Katriel, Journal of Molecular Structure (Theochem) 547, 1
(2001)

\bibitem{Luo2006} M. Luo, Z.B. Li, C.G. Bao, Phys. Rev. A75, 043609 (2007)

\bibitem{Pashov} A. Pashov, et al., Phys. Rev. A 72, 062505 (2005)

\bibitem{NW} It is recalled that the interactions of both the A and B
species are dominated by the spin-independent part, which are both
repulsive. Therefore, we can assume that $\phi _{A}$ and $\phi _{B}$ would
depend on the particle number and $\omega $\ in a similar way. \
Accordingly, $\int d\mathbf{r}|\phi _{A}|^{2}|\phi _{B}|^{2}$ and $\int d%
\mathbf{r}|\phi _{A}|^{4}$ would have similar dependences. It is well known
that the latter is approximately $\propto \omega ^{6/5}N^{-3/5}$. It is
reasonable to suggest that this dependence holds also for the former. This
suggestion has been supported by numerical results.
\end{thebibliography}
\end{document}